\documentclass{aastex631}
\usepackage[T1]{fontenc}        % For typing Polish ogonek symbols (\c{} - A. Kepa co-author name)

\usepackage{graphicx} % Required for inserting images

\graphicspath{{./}{figures/}}

\usepackage{hyperref}

\shorttitle{High-Resolution Solar X-ray Spectroscopy}
\shortauthors{Phillips et al.}

\begin{document}

\title{High-Resolution Solar X-ray Spectroscopy from Archived Solar Maximum Mission Data}

\correspondingauthor{K. J. H. Phillips}
\email{kennethjhphillips@yahoo.com}

\author[0000-0002-3790-990X]{K. J. H. Phillips}\affiliation{Scientific Associate, Earth Sciences Department, Natural History Museum, Cromwell Road, London SW7 5BD, UK}
\email{kennethjhphillips@yahoo.com}

\author[0000-0001-8428-4626]{B. Sylwester}
\affiliation{Space Research Centre, Polish Academy of Sciences (CBK PAN), Warsaw, Bartycka 18A, Poland}
\email{bs@cbk.pan.wroc.pl}

\author[0000-0002-8060-0043]{J. Sylwester}
\affiliation{Space Research Centre, Polish Academy of Sciences (CBK PAN), Warsaw, Bartycka 18A, Poland}
\email{js@cbk.pan.wroc.pl}

\date{\today}

\begin{abstract}

Archived high-resolution X-ray spectra in the 13~\AA\ to 22~\AA\ range from the Flat Crystal Spectrometer (FCS), an instrument on the Solar Maximum Mission operating in the 1980s, are analyzed with reference to nonflaring active regions, and to the \ion{Fe}{17} line emission in light of laboratory and atomic data for nearby \ion{Fe}{16} satellites. The satellites allow temperature to be found for these relatively low-temperature spectra, at which more conventional temperature-dependent line ratios are unavailable. By this means, the spectra can be arranged by temperature, showing that the \ion{Fe}{17} lines are evident at temperatures of $<3$~MK. We confirm that the problem of the underintense Fe XVII 3C and 3D lines is not due to resonant scattering, and instead suggest that, for comparison with CHIANTI spectra, the problem may lie with a needed revision of collisional excitation rates. The line ratio 3G/3H is in theory density-dependent but for \ion{Fe}{17} the ratio is in the low-density limit. However, we suggest that spectra taken during the impulsive stage of flares might reveal a departure from this limit and so allow densities to be derived and hence properties of the flaring plasma. Suggestions for the design of future crystal spectrometers are made in the light of the fluorescence background in FCS spectra.  

\end{abstract}

%\maketitle

%% Section 1%%%%%%%%%%%%%%%%%%%%%%%%%%%%%%%%%%%%%%%%%%%%%%%%%%%%%%%%%%%%%%%%%%%%%%%%%%%%%%%%%%%%%%%%%%%%%%%%
\section{Introduction}
\label{sec:Intro}

The Flat Crystal Spectrometer (FCS) on the NASA spacecraft Solar Maximum Mission (SMM) was a scanning X-ray spectrometer that recorded high-resolution spectra in various wavelength X-ray bands from solar flares and nonflaring active regions over the mission lifetime (1980 February~14 to 1989 November~17). Seven channels covered the entire range, 1.4 to  22.4~\AA, with flat diffracting crystals mounted on a rotatable shaft with rotation angles evaluated by high-precision drive--encoder units. An eighth channel detected white-light emission, allowing context sunspot images to be obtained. For channels~1 to 7, X-rays were incident through a grid collimator giving a triangular response with spatial resolution of approximately 14\arcsec\ FWHM in north--south and east--west directions. The X-ray emission was diffracted from each of the crystals, and the diffracted radiation was detected by proportional counters, one for each crystal. The seven channels had crystals of different material, each with its own value of lattice ($2d$) spacing, giving seven wavelength ranges. The crystals were mounted such that, at a particular angle of the crystal shaft, a strong resonance line was simultaneously detected by all crystal--detector combinations. This so-called home position was used for constructing images of active regions and flares by scanning in solar $x$ (E--W) and $y$ (N--S) directions. Further instrumental details are given by \cite{act80} and \cite{phi82}.  

An early failure of one of the two crystal drive units meant that some caution was exercised in obtaining crystal scans from a time soon after SMM launch (1980 February~14), and large-scale scans for only two large flares, on August 25 and November 5, were accomplished shortly after the maximum activity of solar cycle~21. Spacecraft fine pointing was achieved with three attitude control units, but successive failure of each of these units meant that, after 1980 November, fine pointing was not possible until 1984 April when astronauts on the Space Shuttle Repair Mission (STS 41C) replaced the attitude unit. From this time until 1989 December, when SMM re-entered the Earth's atmosphere, the FCS was again able to operate. Unfortunately, some of the channels had ceased to function through failures of the proportional counter detectors, but a large number of channel~1 (KAP crystal: full range 13.1 -- 22.4 \AA) and channel~3 (ADP crystal: 7.3--10.1~\AA) scans continued to be made until 1987 December~20. These included scans from nonflaring active regions, identifiable from low and practically constant X-ray emission over the duration (typically 14~minutes) of the wavelength scans. 

A companion instrument, the Bent Crystal Spectrometer (BCS), observed mostly flare emission in eight narrow spectral ranges including a group of \ion{Ca}{19} lines and nearby satellite lines. As well as the FCS and BCS, several X-ray spectrometers have been flown on other solar missions, in particular the  P78-1 spacecraft (operational 1979--1985: \cite{mck80,dos83}) covering the 8--23~\AA\ range during flares, had a similar spectral resolution to the FCS. Other instruments with high resolution include the Bragg Crystal Spectrometer (limited spectral ranges around the resonance lines of S, Ca, and Fe) on Yohkoh (operational 1991--2001: \cite{cul91}), and RESIK (REntgenovsky Spektrometr s Izognutymi Kristalami) with spectral range 3.3--6.1~\AA\ on Coronas-F (operational 2001--2003: \cite{jsyl05}). Rocket-borne spectrometers include the high-resolution instrument reported by \cite{par73,par75} and more recently the Marshall Grazing Incidence X-ray Spectrometere (MaGIX) instrument \citep{sav23}. Apart from these, there has been no space-borne solar spectrometer having spectral resolution comparable to that of the FCS in the soft X-ray (10--20~\AA) region. Thus, the FCS channels~1, 2, and 3 ranges, which include numerous Fe lines from Ne-like (\ion{Fe}{17}) to Li-like Fe (\ion{Fe}{24}) as well as lines of He-like ions of other elements, have not been observed with high spectral resolution since the 1980s, and so the archived SMM FCS data remain an important resource for evaluating the physical properties of flares and nonflaring active regions through various line ratios. 

Here we discuss results from an extensive analysis of the newly retrieved data, in particular channel~1 spectra obtained from nonflaring active regions. The archived FCS data for the present analysis were obtained from a publicly available NASA web site\footnote{See https://umbra.nascom.nasa.gov/pub/smm/xrp/data/fis/ for documentation and FCS data (fis) files.}. Among items re-visited here are new estimates of background radiation and spectral line profiles using data for channel~1 crystal rocking curves.  Of particular interest in the channel~1 range are the neon-like Fe (\ion{Fe}{17}) lines, which have received much attention not only from previous discussions of FCS spectra but also from recent laboratory spectra. The intensity of the 15.014~\AA\ line in particular disagrees with calculated values, but it is clear from this study and others that this is not due to optical thickness effects, as had been previously thought, but rather to the need for revised estimates of collisional excitation strengths of some \ion{Fe}{17} lines. The role of \ion{Fe}{16} dielectronic satellites in the 15~\AA\ region is discussed in the context of laboratory data and recent calculations.

Although the spectral resolution of the FCS compares favorably with other space-borne spectrometers, there were shortcomings in the instrument's design that resulted in less-than-ideal observations. We suggest  improvements that can be made and are being implemented in an instrument that is at present under construction. 

%% Section 2%%%%%%%%%%%%%%%%%%%%%%%%%%%%%%%%%%%%%%%%%%%%%%%%%%%%%%%%%%%%%%%%%%%%%%%%%%%%%%%%%%%%%%%%%%%%%%%%
\section{Flat Crystal Spectrometer: Instrument and Data}
\label{sec:FCS_Instr}

% Section 2.1
\subsection{Synposis of FCS Data and Data Selection}
\label{sec:Synopsis}

\begin{figure}
\centerline{\includegraphics[width=0.9\textwidth,clip=,angle=0]{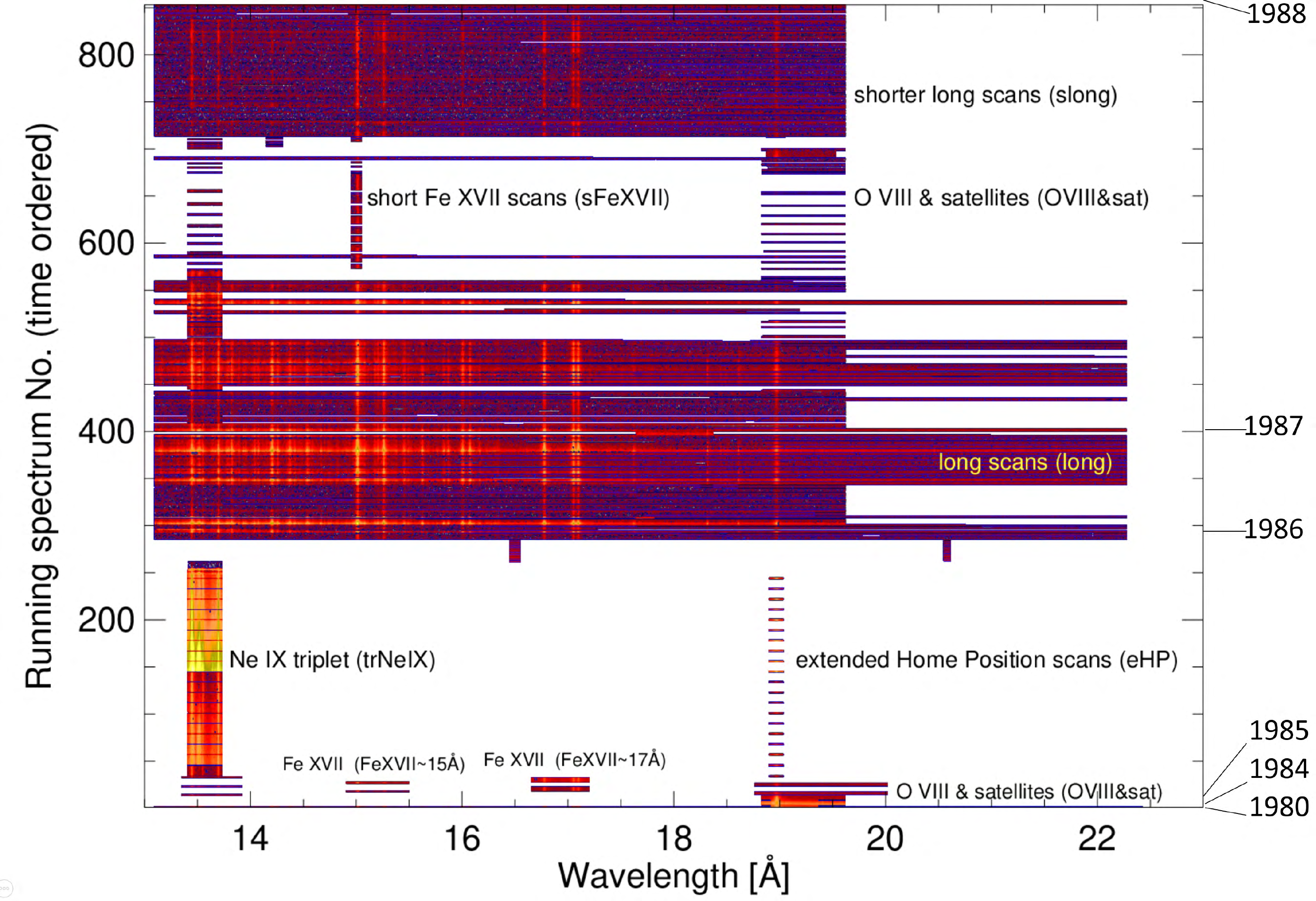}}
\caption{The 853 FCS channel 1 spectral scans, during flares and from nonflaring active regions, are illustrated here in stack form in order of a running spectrum number (indicated on the left-hand vertical axis) and plotted against wavelength (\AA). These spectral scans cover the effective lifetime of FCS~channel~1 (1980 February--1987 December). The colors indicate emission intensity, with high emission colored yellow and lower-level emission colored red or deep red. The strong lines most evident include those of Ne-like Fe (\ion{Fe}{17}) lines (the most intense being at 15.015~\AA, 15.262~\AA, 16.777~\AA, 17.050~\AA, and 17.097~\AA) and the H-like oxygen (\ion{O}{8}) Ly-$\alpha$ line at 18.970~\AA\ (the home position location for channel~1). } \label{fig:fig1}
\end{figure}

Some 853 spectral channel~1 scans are available in the FCS database, including nonflaring active regions and flares with GOES importance up to X2, a range of over a thousand in total emission. We illustrate the extent of the FCS channel~1 spectral dataset in Figure~\ref{fig:fig1} showing color-coded stacks of spectra, plotted in order of a running spectrum number (vertical axis) against wavelength (horizontal axis). An approximate indication of the year of observation is given on the right-hand vertical axis (1980--1987).
%hosted at the Polish Academy of Sciences Space Research Centre (SRC). 
As can be seen from Figure~\ref{fig:fig1}, most channel~1 spectra covered the range 13.1--19.0~\AA, so including the \ion{O}{8} Ly-$\alpha$ line (18.970~\AA), which was the home position for channel~1, and the strong lines between 15~\AA\ and 17~\AA\ due to  Ne-like Fe (\ion{Fe}{17}). Also shown in the figure are much shorter scans covering particular spectral line groups such as the He-like Ne (\ion{Ne}{9}) triplet of lines (13.477--13.698~\AA) and scans over the profile of the \ion{O}{8} Ly-$\alpha$ line. A few spectra included the diagnostically important He-like oxygen (\ion{O}{7} line triplet at 21.602--22.101~\AA), although these lines were always very weak owing to decreased sensitivity at these wavelengths. Lines due to He-like Mg (\ion{Mg}{11}) at 9.1685--9.314~\AA\ occur in second order. 

A total of 165 channel~1 spectra from nonflaring active regions were identified in the 1985--1987 period for which there were extremely small or practically zero changes in the background emission over the scan periods. From these spectra (details of which are given on the web site referenced below), we selected spectra with the strongest lines having a photon count rate of less than $\sim 100~{\rm s}^{-1}$, corresponding to GOES 1--8~\AA\ emission from A4.9 to B8.0. A complete list of spectra with details of the emitting region's heliographic location and GOES X-ray importance is available on the web site (\verb!http://www.cbk.pan.wroc.pl/js/APJ/Final_list_of_AR_spectra.txt!) 

%Fig%%%%%%%%%%%%%%%%%%%%%%%%%%%%%%%%%%%%%%%%%%%%%%%%%%%%%%%%%%%%%%%%%%%%%%%%%%%%%%%%%%%%%%%%%%%%%%%%%%%%%%%%%%
%% Figure 1
%Fig%%%%%%%%%%%%%%%%%%%%%%%%%%%%%%%%%%%%%%%%%%%%%%%%%%%%%%%%%%%%%%%%%%%%%%%%%%%%%%%%%%%%%%%%%%%%%%%%%%%%%%%%%%

FCS spectra from  all channels had a background of instrumental origin, mostly due to fluorescence emission from the FCS crystals arising from solar X-ray emission incident on the crystals as well as the FCS structure including the thermal shields. A smaller contribution was made by energetic particles. Each of the seven detectors viewed the fluorescence from all the crystals, the total emission forming a significant background, or ``pedestal'', to the spectra in all channels, in particular those from channel~1 discussed here. This was mitigated by the fact that for this analysis the spectra were from nonflaring active regions with GOES B-level emission or less, so reducing the amount of fluorescence. 

% Section 2.2
\subsection{Background Subtraction}
\label{sec:Background}

\begin{figure}
\centerline{\includegraphics[width=0.9\textwidth,clip=,angle=0]{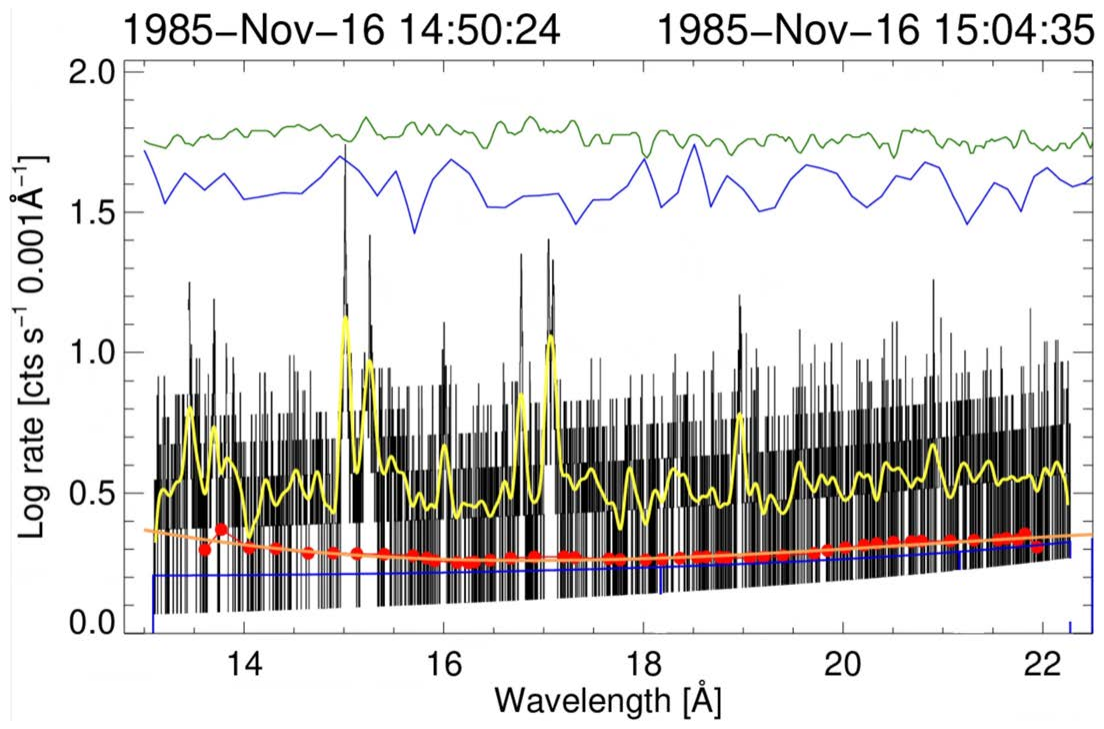}}
\caption{Procedure for estimating background (pedestal) level for a single nonflaring active region FCS channel~1 spectrum (14.5-minute duration starting 1985 November~16 at 14:50:24~UT). Black histogram: observed spectrum plotted logarithmically (units of photon counts~s$^{-1}$~0.001\AA$^{-1}$). Yellow curve: smoothed observed spectrum with a box-car average having a width 0.005~\AA. Red dots: minima determined using an automated IDL routine, which are minima in the yellow curve subtracted by 0.2 in the logarithm. Orange curve: Third-order polynomial to fit the red dots. Lower blue curve: preliminary determination pf the pedestal by eye. Green curve: GOES 1--8~\AA\ emission plotted logarithmically to illustrate the absence of large changes with time (and therefore wavelength) during the FCS scan. The vertical scale of this curve is arbitrary (as drawn it corresponds to GOES level A5.8). Upper blue curve: total flux in \ion{Ca}{19} BCS channel~1.} \label{fig:fig2}
\end{figure}

For the  analysis of channel~1 spectra, an accurate subtraction of the pedestal emission was first necessary for analysis of the observed line emission. We analyzed each of the 165 nonflaring active region spectra (using Interactive Data Language, IDL) by first plotting the raw spectrum in units of photon counts~s$^{-1}$ 0.001~\AA$^{-1}$. Figure~\ref{fig:fig2} is an example, the black histogram being the raw spectrum. To enable the spectral lines to be better distinguished, the raw spectrum was smoothed with a box-car average having a width 0.005~\AA\ (yellow curve in the figure). As the smoothing process raises the pedestal level, we took the true pedestal level to be lower than the minima in the yellow curve by 0.2 in the logarithm. These minima are shown by the red dots in the figure, and a third-order polynomial curve fitted to these points results in the orange line. We defined this to be the pedestal level, which was then subtracted from this spectrum. As a check on any significant time variability of this pedestal level over the scan period, we plotted with time the GOES 1--8~\AA\ total X-ray emission (green curve) and the total flux recorded by the Bent Crystal Spectrometer in its channel~1, viewing the \ion{Ca}{19} spectrum (upper blue curve). The spectrum in Figure~\ref{fig:fig2} was taken on 1985 November~16, and is one of 10 nonflaring active region spectra that included the extended range, to 22.26~\AA. We found that this automated method for the determination of the pedestal level was an improvement over a method used in a preliminary analysis in which the background was estimated by eye (this is shown as the lower blue curve in the figure).

%Fig%%%%%%%%%%%%%%%%%%%%%%%%%%%%%%%%%%%%%%%%%%%%%%%%%%%%%%%%%%%%%%%%%%%%%%%%%%%%%%%%%%%%%%%%%%%%%%%%%%%%%%%%%%
% Figure 2
%Fig%%%%%%%%%%%%%%%%%%%%%%%%%%%%%%%%%%%%%%%%%%%%%%%%%%%%%%%%%%%%%%%%%%%%%%%%%%%%%%%%%%%%%%%%%%%%%%%%%%%%%%%%%%

\begin{figure}
\centerline{\includegraphics[width=0.90\textwidth,clip=,angle=0]{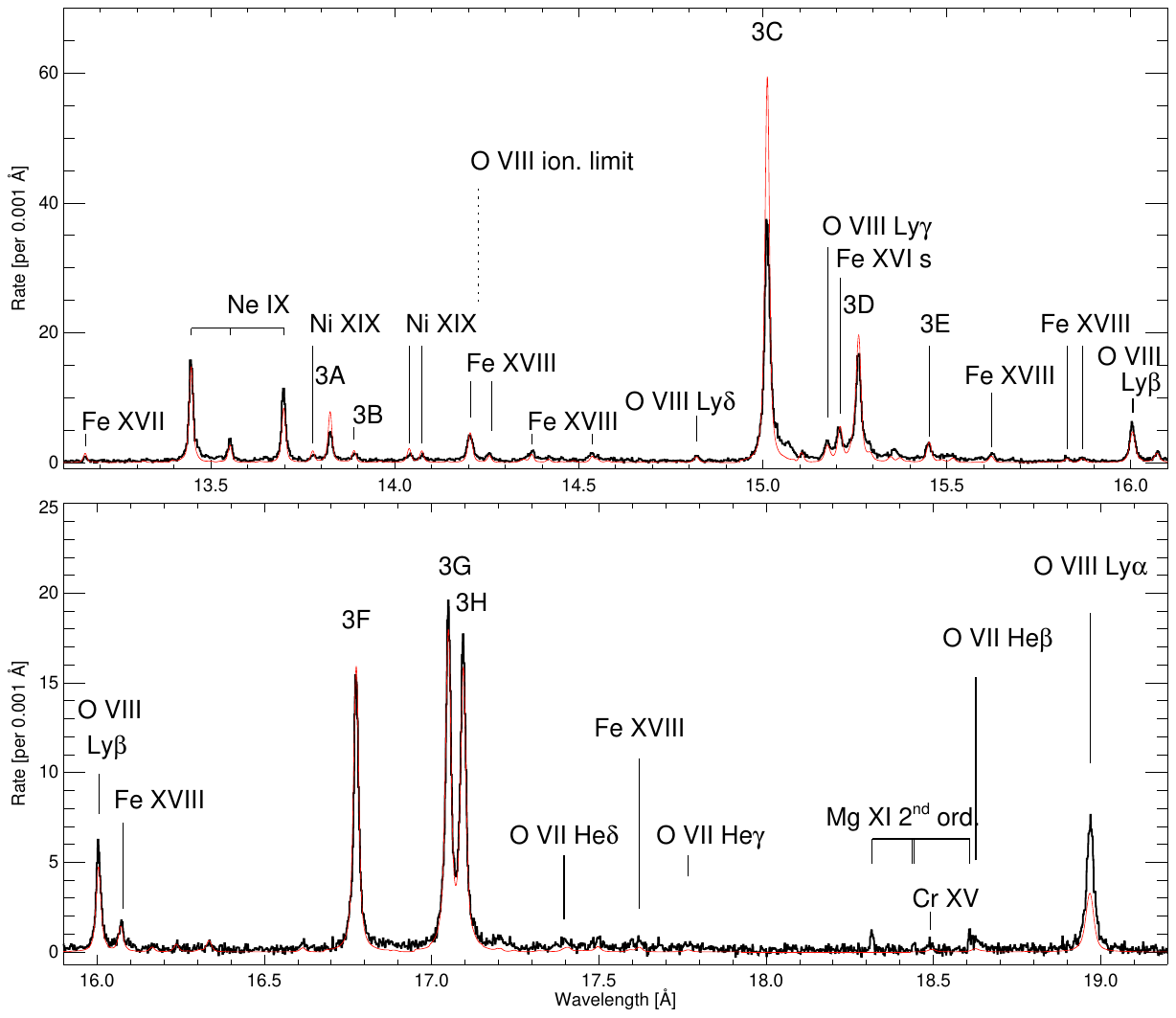}}
\caption{The FCS average spectrum for 165 nonflaring active region spectra (in black) displayed in two sections (13.1--15.8~\AA, 15.7--19.6~\AA). A comparison CHIANTI v.~11 (with input coronal abundances of  \cite{fel92b} and ion fractions of \cite{bry06}) with \ion{Fe}{16} satellites is shown as the red curve, calculated for temperature $T = 3.80$~MK and volume emission measure = $8.85 \times 10^{45}$~cm$^{-3}$. Principal lines are identified as in the last three columns of Table~\ref{tab:line_ids} (\ion{Fe}{17} lines with the notation of \cite{par73}). The ionization limit of H-like oxygen is also shown. } \label{fig:fig3}
\end{figure}

An average of the 165 spectra in the 13.1--19.5~\AA\ range is shown in Figure~\ref{fig:fig3} with pedestal subtracted (black histogram). The red curve is a theoretical spectrum from the CHIANTI atomic database and code (version 11: \cite{duFr24}) discussed in the next section. Chief spectral lines are indicated by the emitting ions, with those of \ion{Fe}{17} identified by letters and numbers (3A to 3H) given by \cite{par73} in a discussion of a high-resolution spectrum from a rocket-borne crystal spectrometer. Details of individual lines (wavelengths, intensities, corresponding atomic transitions, and line notation) are given in Table~\ref{tab:line_ids}. Disagreements of this average spectrum with the CHIANTI theory are evident for some of the \ion{Fe}{17} lines (3A, 3C, 3D), which are discussed below. There is also a factor-of-two mismatch between the observed and CHIANTI spectrum for the \ion{O}{8} Ly-$\alpha$ line at 18.970~\AA\ although not for other hydrogen-like oxygen lines Ly-$\beta$, Ly-$\gamma$, and Ly-$\delta$. This is possibly attributable to an increasingly uncertain calculated sensitivity of FCS channel~1 at longer wavelengths for which the sensitivity is steeply falling.

%Fig%%%%%%%%%%%%%%%%%%%%%%%%%%%%%%%%%%%%%%%%%%%%%%%%%%%%%%%%%%%%%%%%%%%%%%%%%%%%%%%%%%%%%%%%%%%%%%%%%%%%%%%%%%
% Figure 3
%Fig%%%%%%%%%%%%%%%%%%%%%%%%%%%%%%%%%%%%%%%%%%%%%%%%%%%%%%%%%%%%%%%%%%%%%%%%%%%%%%%%%%%%%%%%%%%%%%%%%%%%%%%%%%

% Table 1  - Nonflaring AR line list with lines evident in Fig. 7 (CHIANTI and our measured wvls compared).
\begin{deluxetable*}{ccrlll}
\tabletypesize{\small}
\tablecaption{Nonflaring Active Region FCS Channel~1 Line List 13.1--19.0~\AA. \label{tab:line_ids} }
\tablewidth{0pt}
\tablehead{\colhead{CHIANTI line }  & \colhead{Obs. wavelength} & \colhead{Line intensity$^3$} & \colhead{Ion} & \colhead{Transition$^4$}  &\colhead{Notation$^5$}\\
\colhead{wavelength (\AA)$^1$} & \colhead{(uncertainty)$^2$} & \colhead{} }
\startdata   
13.1594 & 13.1598 (30) & 168.9    & \ion{Fe}{17} & $2s^2 2p^6 \, ^1S_0 - 2s 2p^6 3d \, ^1D_2 $  \\
13.4473 & 13.4493 (17) & 1512.4 & \ion{Ne}{9} & $1s^2 \, ^1S_0 - 1s2p \, ^1P_1$ & $w$\\
13.5531 & 13.5506 (20) & 404.2 & \ion{Ne}{9} & $1s^2 \, ^1S_0 - 1s2p \, ^3P_1$ & $y$\\
13.6990 & 13.6983 (18) &  1113.1 & \ion{Ne}{9} & $1s^2 \, ^1S_0 - 1s2s \, ^3S_1$ & $z$ \\
13.7777 & 13.7781 (30) & 210.0 & \ion{Ni}{19} & $2p^6 \, ^1S_0 - 2p^5 3s \, ^1P_1$ & Equiv. of 3F in \ion{Ni}{19}\\
13.8250 & 13.8252 (19) & 635.4 & \ion{Fe}{17} & $2s^2 2p^6\,^1S_0 - 2s2p^63p \, ^1P_1$ & 3A\\
13.8900 & 13.8933 (26) & 354.7 & \ion{Fe}{17} & $2s^2 2p^6\,^1S_0 - 2s2p^63p \, ^3P_1$ & 3B\\
14.0398 & 14.0450 (30) & 360.0 & \ion{Ni}{19} & $2p^6 \, ^1S_0 - 2p^5 3s \, ^3P_1$ & Equiv. of 3G in \ion{Ni}{19} \\
14.0741 & 14.0760 (30) & 350.0 & \ion{Ni}{19} & $2p^6 \, ^1S_0 - 2p^5 3s \, ^3P_2$ & Equiv. of 3H in \ion{Ni}{19} \\
14.2057 & 14.2084 (24) & 833.7 & \ion{Fe}{18} & $2s^2 2p^5 \, ^2P_{3/2} - 2s^2 2p^4 (^1D) 3d \,\, ^2D_{5/2}$ \\
14.2580 & 14.2620 (26) & 155.0 & \ion{Fe}{18} & $2s^2 2p^5 \, ^2P_{3/2} - 2s^2 2p^4 (^1D) 3d\,\,^2S_{1/2}, ^2F_{5/2}$ & \\
14.3730 & 14.3745 (36) & 704.4 & \ion{Fe}{18} & $2s^2 2p^5 \, ^2P_{3/2} - 2s^2 2p^4 (^3P) 3d\,\, ^2D_{5/2}$ \\
14.5370 & 14.5401 (38) & 718.5 & \ion{Fe}{18} & $2s^2 2p^5 \, ^2P_{3/2} - 2s^2 2p^4 (^3P) 3d\,\, ^2F_{5/2}$ \\
14.8206 & 14.8214 (34) & 517.5 & \ion{O}{8}   & $1s ^2S_{1/2} - 5p\, ^2P_{1/2,3/2}$ &  Ly-$\delta$ \\
15.0130 & 15.0140 (16) & 7763.6& \ion{Fe}{17} & $2s^2 2p^6\,^1S_0 - 2s^2 2p^5\,3d \, ^1P_1$ & 3C \\
15.1050 & 15.1100 (30) & 470.0 & \ion{Fe}{16} & $2s^2 2p^6\,3s ^2S_{1/2} - 2s^2 2p^5 3s 3d \,^2P_{3/2}$ & Diel. sat. (``A'')$^6$ \\
15.1762 & 15.1752 (22) & 980.2 & \ion{O}{8} & $1s ^2S_{1/2} - 4p ^2P_{1/2,3/2}$ &  Ly-$\gamma$ \\
15.2078 & 15.2080 (28) & 1220.0 & \ion{Fe}{16} & $2s^2 2p^6\,3s ^2S_{1/2} - 2s^2 2p^5 3s 3d \,\,^2P_{1/2}$ & Diel. sat. (``B'')$^6$ \\
15.2620 & 15.2611 (14) & 3782.7 & \ion{Fe}{17} & $2s^2 2p^6\,3d ^1S_0 - 2s^2 2p^5\,3d \, ^3D_1$ & 3D \\
15.2620 & Blended with 3D & ----- & \ion{Fe}{16} & $2s^2 2p^6\,3s\, ^2S_{1/2} - 2s^2 2p^5 3s3d\, ^2D_{3/2}$ & Diel. sat. (``C'')$^6$ \\
15.3550 & 15.3560 (20) & 1050.0 & \ion{Fe}{16}  & $2s^2 2p^6\,3d\, ^2D_{3/2} - 2s^2 2p^5 3d^2\, ^2F_{5/2}$ & Diel. sat. (``$\alpha$'')$^6$ \\
15.4520 & 15.4543 (26) & 1216.6 & \ion{Fe}{17} & $2s^2 2p^6\,^1S_0 - 2s^2 2p^5\,3d \, ^3P_1$ & 3E \\
15.6221 & 15.6240 (29) & 766.3 & \ion{Fe}{18} & $2s^2 2p^5 \, ^2P_{3/2} - 2s^2 2p^4\, (^1D) 3s \, ^2D_{5/2}$ \\
15.8280 & 15.8300 (43) & 855.0 & \ion{Fe}{18} & $2s^2 2p^5 \, ^2P_{3/2} - 2s^2 2p^4\, (^3P) 3s \, ^4P_{3/2}$ \\
15.8700 & 15.8703 (37) & 886.2 & \ion{Fe}{18} & $2s^2 2p^5 \, ^2P_{1/2} - 2s^2 2p^4\, (^1D) 3s \, ^2D_{3/2}$ \\
16.0059 & 16.0080 (27) & 2648.4 & \ion{O}{8} & $1s ^2S_{1/2} - 3p\, ^2P_{1/2,3/2}$  &  Ly-$\beta$ \\ 
16.0720 & 16.0748 (34) & 1452.2 & \ion{Fe}{18} & $2s^2 2p^5 \, ^2P_{3/2} - 2s^2 2p^4\, (^3P) 3s \, ^4P_{5/2}$ \\
16.7756 & 16.7768 (32) & 8443.7 & \ion{Fe}{17} & $2s^2 2p^6\,^1S_0 - 2s^2 2p^5\,3s \, ^1P_1$ & 3F \\
17.0510 & 17.0504 (14) & 10891.9 & \ion{Fe}{17} & $2s^2 2p^6\,^1S_0 - 2s^2 2p^5\,3s \, ^3P_1$ & 3G \\
17.0960 & 17.0957 (14) & 10197.7 & \ion{Fe}{17} & $2s^2 2p^6\,^1S_0 - 2s^2 2p^5\,3s \, ^3P_2$ & 3H \\ 
17.3960 & 17.3950 (46) & 3594.2 & \ion{O}{7} & $1s^2\,^1S_{0} - 1s5p\,^1P_{1}$ &  He$\delta$ \\
17.6218 & 17.6092 (46) & 4636.6 & \ion{Fe}{18} & $2s2p^6 \, ^2S_{1/2} - 2s^2 2p^4\, (^1D) 3p \, ^2P_{3/2}$ \\
17.7680 & 17.7724 (36) & 3433.3 & \ion{O}{7} & $1s^2\,^1S_{0} - 1s4p\,^1P_{1}$ &  He$\gamma$ \\
9.1687  & 18.3166 (29)  & 39000 & \ion{Mg}{11} & $1s^2\,^1S_0 - 1s2p\,^1P_1$ & $w$ (in 2nd order)$^7$ \\
9.2312  & 18.4459 (30)  & 19000   & \ion{Mg}{11} & $1s^2\,^1S_0 - 1s2p\,^3P_1$ & $y$ (in 2nd order) \\
18.4969 & 18.4902 (39) & 3652.4 & \ion{Cr}{15} & $2s^2 2p^6\,^1S_0 - 2s^2 2p^5\,3d \, ^1P_1$ & Equiv. of 3C in \ion{Cr}{15} \\
9.3143  & 18.6065 (30)  & 40000 & \ion{Mg}{11} & $1s^2\,^1S_0 - 1s2s\,^3S_1$ & $z$ (in 2nd order) \\
18.6270 & 18.6450 (33) & 4396.1 & \ion{O}{7} & $1s^2\,^1S_{0} - 1s3p\,^1P_{1}$ &  He$\beta$ \\
18.9689 & 18.9696 (22) & 22210.0 & \ion{O}{8} & $1s\,^2S_{1/2} - 2p\,^2P_{1/2,3/2}$  &  Ly-$\alpha$ \\
\enddata
\tablenotetext{1}{From CHIANTI v.~11 \citep{duFr24} and \cite{delz24}. For \ion{Fe}{15} and \ion{Fe}{16} dielectronic satellite notation and wavelengths, see \cite{bro06}. Cols. 1, 5: line identifications and predicted wavelengths are from CHIANTI v.11. }
\tablenotetext{2}{Statistical uncertainties in parentheses, given in units of 0.0001~\AA.}
\tablenotetext{3}{From average of 165 spectra (Figure~\ref{fig:fig3}). Units of photon counts cm$^{-2}$ s$^{-1}$.}
\tablenotetext{4}{\ion{Fe}{17} and \ion{Ni}{19} transitions in $LS$-coupling notation. See \cite{kel87} for equivalent $jj$-coupling notation.}
\tablenotetext{5}{\ion{Fe}{17} line notation from \cite{par73}. \ion{Ne}{9}, \ion{Mg}{11} line notation from \cite{gab72}.}
\tablenotetext{6}{Transitions and wavelengths from \cite{delz24}. See \cite{bro06} for their identifications of satellite features ``A'', ``B'', ``C'', and ``$\alpha$'' (line ``$\alpha$'' was identified by \cite{bro06} with an \ion{Fe}{15} transition).}
\tablenotetext{7}{Intensities of second-order \ion{Mg}{11} lines estimated from \cite{bur76}.}
\end{deluxetable*}

% Subsection 2.3
\subsection{Spectral resolution and wavelength precision for FCS channel 1}
\label{sec:line_profile}

The spectral resolution of FCS channel~1 compares well with any solar X-ray spectrometer flown to date, as indicated by the narrow rocking curve widths for the KAP crystal used for this channel. Pre-launch measurements of the KAP crystal characteristics are available, but improvements on these data including the wavelength dependence of the rocking curve width and integrated reflectivity were made using the widely used X-ray Optics Software Toolkit (XOP: for version 2.1, see \cite{delR04}). The rocking curve full width half maximum (FWHM) varies by more than a factor of 3 over the wavelength range of the spectral scans discussed here, with the XOP values being slightly less than the pre-launch estimates. As deduced from the XOP data, the shape of the rocking curve is close to a Voigt profile with a slightly asymmetrical central part. This profile was convolved with a CHIANTI spectrum having a temperature 3.80~MK. When multiplied by the pre-launch effective area for channel~1, this curve is shown in Figure~\ref{fig:fig3} and illustrates the close agreement with the observed averaged spectrum for the 165 nonflaring active regions. Exceptions were noted in Section~\ref{sec:Background}, but in addition the CHIANTI spectrum convolved with the instrumental curve initially gave too low a minimum between the two closely spaced \ion{Fe}{17} 3G and 3H lines. This was rectified by adding a Gaussian profile with FWHM of $0.0045 \pm 0.0003$~\AA\ constant with wavelength to all observed lines (in the calculation with the XOP software, a reference wavelength is needed which we took to be 12.02~\AA, below the lower bound of FCS channel~1). The addition of this component may be physically explained by slight imperfections of the flight crystal or possibly random turbulent velocities in the nonflaring active regions. The line intensities as given in Table~\ref{tab:line_ids} were deduced from the integral over the line profiles. Those for the second-order \ion{Mg}{11} were estimated using second-order KAP reflectivities from Figure~7 of \cite{bur76}.

The entire FCS channel~1 range had a very high spectral precision because of the high-precision Baldwin drive units allowing spectra to be obtained knowing the best estimate of the wavelength of the home position line \citep{act80}. In the case of channel~1, this was the Ly-$\alpha$ doublet of \ion{O}{8} from \cite{eri77}, the combined wavelength of the two fine-structure components being 18.9689~\AA. A measure of the FCS wavelength precision is provided by the \ion{Ne}{9} resonance ($w$) line: the observed wavelength is 13.4473~\AA, the theoretical wavelength 13.4471~\AA, based on the Dirac--Fock calculations of \cite{hat82}, so a difference of 0.0002~\AA. For the \ion{Mg}{11} seen in second order, the observed wavelength is 9.1687~\AA, the \cite{hat82} wavelength 9.1685~\AA, or a difference of 0.0002~\AA. We may conclude that the observed wavelength precision indicated in Table~\ref{tab:line_ids} is about 0.0002~\AA.

% Subsection 2.4
\subsection{Temperature Assignment}
\label{sec:T_assign}

In the comparison of the observed averaged nonflaring active region spectrum with that calculated from the CHIANTI atomic code (Figure~\ref{fig:fig3}), we used the latest release (version 11: \cite{duFr24}) together with recent calculations of dielectronic satellites of \ion{Fe}{16} given by \cite{delz24} and G. Del Zanna (2025, private communication). The line intensities are temperature-dependent because of the collisional excitation mechanisms and the fractional abundance of ions from ionization equilibrium calculations \citep{bry06}, the emitting plasma for a nonflaring active region being assumed to be in steady state. Although no specific flares could be identified in any of these spectra, the GOES 1--8~\AA\ emission often showed very small variations in X-ray emission that can be ascribed to low level activity like mini-flares. A temperature from line ratios in a particular spectrum will reflect the amount of this nonflaring activity, and so a temperature characterizing each spectrum is therefore desirable. The ratio of the two GOES (0.5--4~\AA\ and 1--8~\AA) channels has been used in our previous analyses of SMM BCS \ion{Ca}{19} spectra \citep{jsyl22,bsyl23}. Its use is always subject to the proviso that GOES, being a full-Sun instrument, viewed X-ray emission from all active regions on the disk if present at the time of observations. This is less of a concern for the present analysis dealing with FCS spectra taken in the 1985--1987 period, as activity then (shortly after the minimum between Cycles 21 and 22) was low while an active region observed was generally the only significant one on the visible solar disk at any time. We therefore derived this temperature, denoted by $T_{\rm G}$, for comparison with those from line ratios. One such ratio that we considered involves the sum of the intercombination and forbidden lines ($y$ and $z$) to the resonance line $w$ in the \ion{Ne}{9} line group (13.447--13.699~\AA) included in channel~1, the so-called $G$ ratio \citep{gab69}). However, the temperature dependence is too slight to be of use for the spectra analyzed. Another possible line ratio is that of one of the more intense \ion{Fe}{17} lines to an \ion{Fe}{18} line, the most intense of which is at 14.206~\AA. Unfortunately, the \ion{Fe}{18} was too weak for spectra at times of very low activity (and presumably low temperature), these being of particular interest for the study of \ion{Fe}{16} dielectronic satellites that occur in the range of the \ion{Fe}{17} lines. 

\begin{figure}
\centerline{\includegraphics[width=0.8\textwidth,clip=,angle=90]{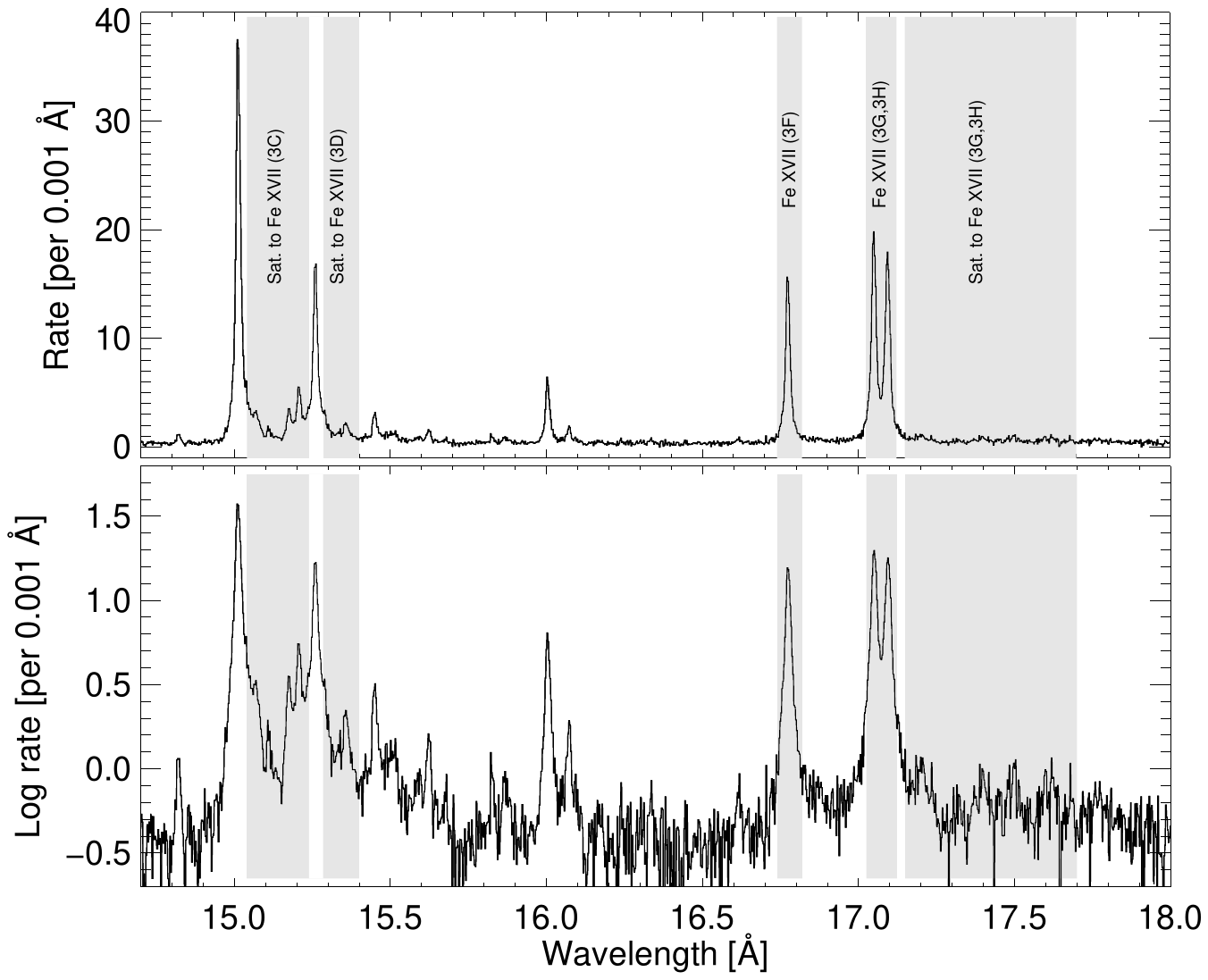} }
\caption{Upper panel: Average nonflaring active region spectrum (linear scale) showing bands used for estimating temperatures. For the temperature ${T_{15{\rm sat}}}$, the ratio of Fe~XVI satellites in the bands on the long-wavelength side of  lines 3C and 3D (marked in gray) to Fe~XVII line 3F (16.776~\AA). For the temperature ${T_{17{\rm sat}}}$, the ratio of Fe~XVI satellites in the bands on the long-wavelength side of \ion{Fe}{17} lines 3G and 3H (marked in gray) to Fe~XVII line 3F. Lower panel: Same as upper panel but plotted on a logarithmic scale.
\label{fig:fig4} }
\end{figure}

\begin{figure}
\centerline{\includegraphics[width=1.0\textwidth,clip=,angle=0]{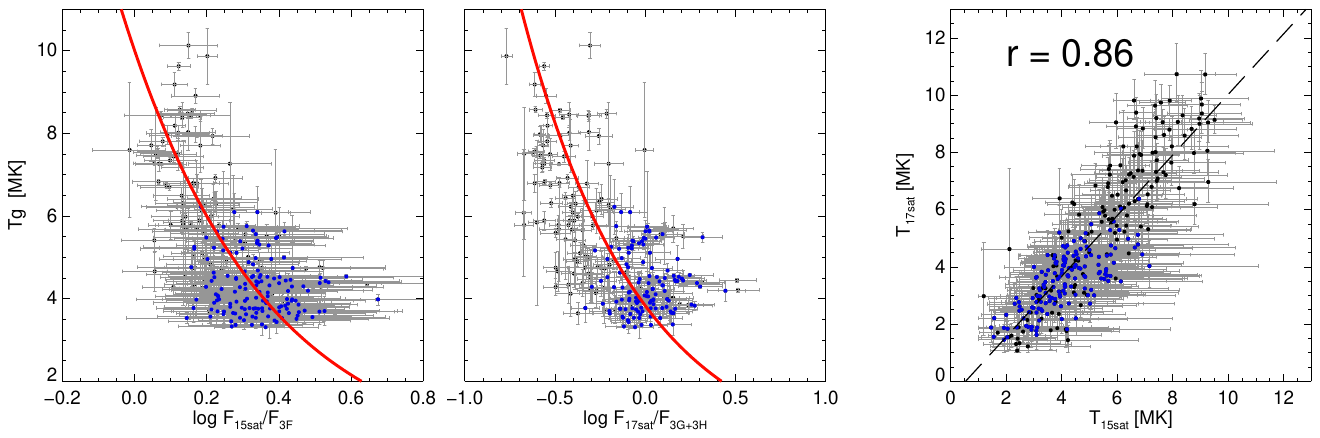} }
\caption{Left: GOES temperature $T_{\rm  G}$ plotted against the intensity ratio $F_{15{\rm sat}}/F_{3F}$ for both flare (in gray) and nonflaring active regions (blue points). The error bars represent statistical uncertainties. The red line is the best-fit $T^{-0.9}$ dependence. Middle: As for the left panel but with the intensity ratio $F_{17{\rm sat}}/F_{3G+3H}$ on the $x$-axis. The red line is the best-fit $T^{-1.5}$ dependence. Right: Plot of $T_{15{\rm sat}}$ against $T_{17{\rm sat}}$ showing the strong correlation (coefficient $r = 0.86$).
\label{fig:fig5} }
\end{figure}

As mentioned in the recent work by \cite{delz24}, the numerous \ion{Fe}{16} dielectronic satellites offer another means of estimating temperature for nonflaring active regions. Wavelengths and intensity factors for these satellites were calculated by \cite{cor94} and \cite{phi97}, but here we use the data of \cite{delz24} which also includes inner-shell excitation. Many thousands of satellites occur in the region (15~\AA\ -- 17~\AA) of the \ion{Fe}{17} lines, nearly all very weak but their cumulative effect is important.  Accordingly, using the \cite{delz24} data, we selected the line ratio of all \ion{Fe}{16} satellites on the long-wavelength side of the \ion{Fe}{17} lines 3C and 3D (intensity denoted by $F_{15{\rm sat}}$) and another group of satellites on the long-wavelength side of lines 3G and 3H (intensity $F_{17{\rm sat}}$) to the \ion{Fe}{17} line 3F. Line 3F is unblended, unlike lines 3C and 3D near to which are lines apparent at lower temperatures.  Figure~\ref{fig:fig4} shows two wavelength bands (marked in gray) that include \ion{Fe}{16} satellites near the \ion{Fe}{17} lines 3C and 3D and a single band for satellites on the long-wavelength side of \ion{Fe}{17} lines 3F, 3G, and 3H. The band marked ``Sat. to Fe XVII (3C)''  in Figure~\ref{fig:fig4} also includes the \ion{O}{8} Ly-$\gamma$ line at 15.1752~\AA\ (observed wavelength), but we neglected the contribution of this relatively weak line. The temperatures derived from these two ratios, $T_{15{\rm sat}}$ and $T_{17{\rm sat}}$, are highly correlated with each other, with a Spearman's correlation coefficient of $r = 0.86$ using both nonflaring active region and flare spectra, a total of 307 points (Figure~\ref{fig:fig5}, right panel). For nonflaring active region spectra alone (158 spectra out of 165 for which both temperature values were available), the correlation coefficient is 0.71. We also compared the intensity ratio $F_{15{\rm sat}}/F_{3F}$ and $F_{17{\rm sat}}/F_{3G + 3H}$ for all 853 spectra that includes flares (black dots with error bars) and nonflaring active regions (blue dots) with the GOES temperature $T_{\rm G}$ in Figure~\ref{fig:fig5} (left and center panels). We checked that all these spectra showed little or no time variations in the 14-minute scan period. The red curves in these plots are best-fit $T^{-n}$ dependence with $n=0.9$ ($F_{15{\rm sat}}/F_{3F}$) and $n=1.5$ ($F_{17{\rm sat}}/F_{3G + 3H}$). For the purposes of this study, we take the mean, $(T_{15{\rm sat}} + T_{17{\rm sat}})/2$, called henceforth $T_{1517}$, to be the temperature characterizing the active region spectra. Departures of a few points from the red curves are attributable to the fact that, for these few occasions, GOES observed X-ray emission not only from the active region as observed by the FCS but also from emission outside the FCS field of view.

%Fig%%%%%%%%%%%%%%%%%%%%%%%%%%%%%%%%%%%%%%%%%%%%%%%%%%%%%%%%%%%%%%%%%%%%%%%%%%%%%%%%%%%%%%%%%%%%%%%%%%%%%%%%%%
% Figure 4
% \ion{Fe}{17}
%Fig%%%%%%%%%%%%%%%%%%%%%%%%%%%%%%%%%%%%%%%%%%%%%%%%%%%%%%%%%%%%%%%%%%%%%%%%%%%%%%%%%%%%%%%%%%%%%%%%%%%%%%%%%%

%Fig%%%%%%%%%%%%%%%%%%%%%%%%%%%%%%%%%%%%%%%%%%%%%%%%%%%%%%%%%%%%%%%%%%%%%%%%%%%%%%%%%%%%%%%%%%%%%%%%%%%%%%%%%%
% Figure 5
%Fig%%%%%%%%%%%%%%%%%%%%%%%%%%%%%%%%%%%%%%%%%%%%%%%%%%%%%%%%%%%%%%%%%%%%%%%%%%%%%%%%%%%%%%%%%%%%%%%%%%%%%%%%%%

%% Section 3%%%%%%%%%%%%%%%%%%%%%%%%%%%%%%%%%%%%%%%%%%%%%%%%%%%%%%%%%%%%%%%%%%%%%%%%%%%%%%%%%%%%%%%%%%%%%%%%
% Fe XVII lines in AR spectra
\section{Ionized Iron Lines in FCS Active Region Spectra}
\label{sec:FeXVII_lines}

% Subsection 3.1
\subsection{Fe XVII Lines}
\label{FeXVI}

As is clear from Figure~\ref{fig:fig3}, lines of \ion{Fe}{17} between 15~\AA\ and 17~\AA\ dominate the channel~1 spectrum. These lines in solar spectra have been widely commented on (see \cite{par73} and references therein) and have been observed in non-solar sources such as Capella \citep{beh01,phi01}. Because of problems associated particularly with the apparent under-intensity of the 3C line (15.014~\AA) line (e.g. \cite{ber12,kuh22}), studies with laboratory devices, particularly Electron Beam Ion Traps (EBIT), have tried to elucidate the origins of these problems. Here we point out that, although much light has been shed on the problem, the FCS spectra analyzed here show that there remains an important discrepancy when comparison is made with CHIANTI calculations. 

\begin{figure}
\centerline{\includegraphics[width=0.9\textwidth,clip=,angle=0]{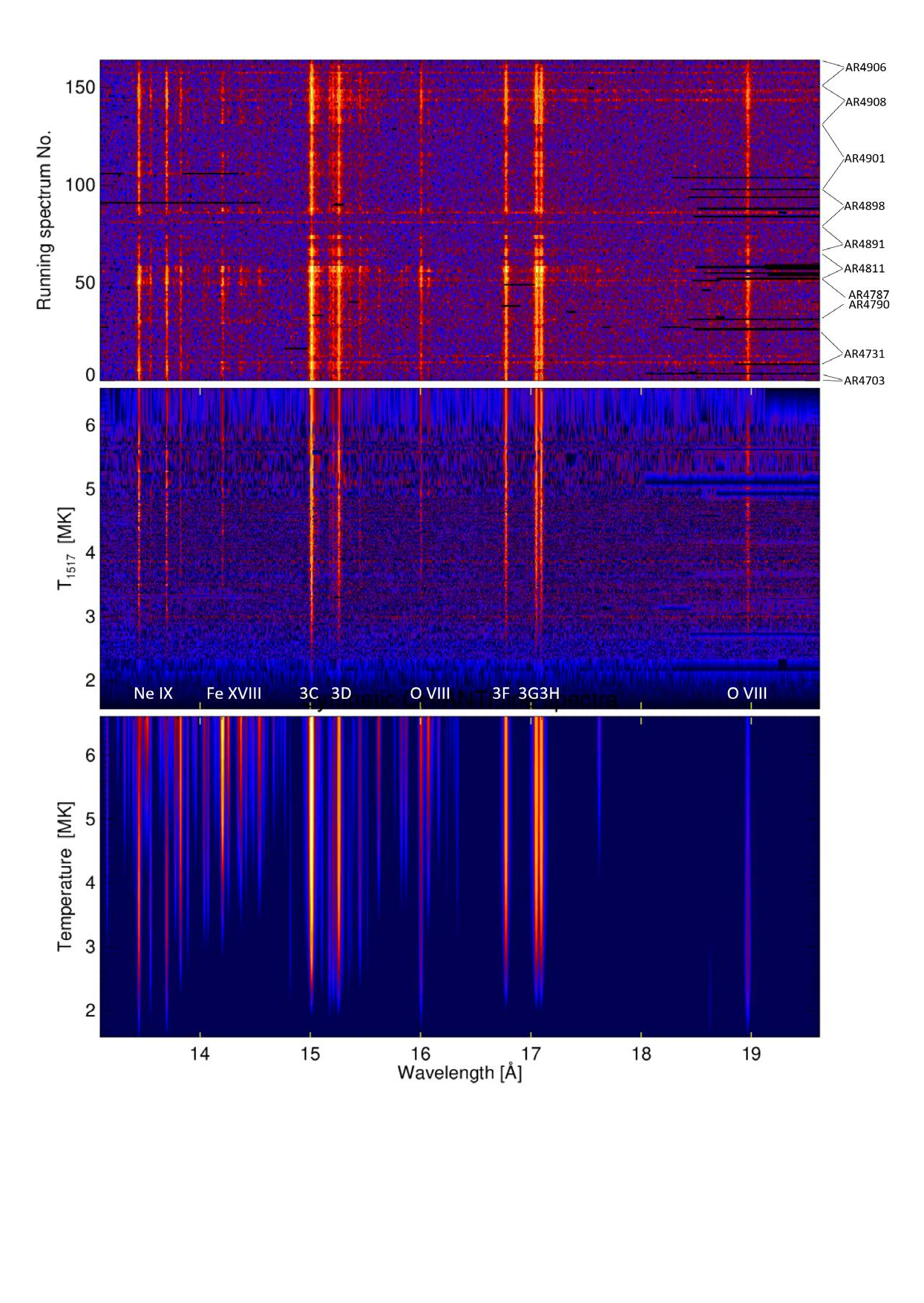}}
\caption{Top: All 165 FCS nonflaring active region spectra, stacked by time (earliest times at bottom); the time range of the NOAA active region number is indicated on the right-hand scale. Middle: The same spectra stacked by temperature $T_{1517}$ (mean of $T_{15{\rm sat}}$ and $T_{17{\rm sat}}$), for unit emission measure, temperature increasing upwards. Bottom: Equivalent CHIANTI v.~11 spectra stacked by temperature over the same range as the middle panel.} \label{fig:fig6}
\end{figure}

Armed with the temperature $T_{1517}$ determined  from \ion{Fe}{16} satellites, we are able to sort the active region spectra according to temperature $T_{1517}$. In Figure~\ref{fig:fig6} all 165 active-region spectra are shown in stack form, with spectra ordered in time (top panel) and temperature $T_{1517}$ (middle panel) after normalization to unit emission measure. Active region numbers from the National Oceanic and Atmospheric Administration (NOAA) scheme are given on the right-hand vertical axis for the top panel. The bottom panel shows CHIANTI spectra over the same temperature range as calculated per unit emission measure. It is seen from the center panel that the principal \ion{Fe}{17} lines in the observed spectra are prominent at all temperatures and even extend to $T_{1517} < 3$~MK for lines 3C--3H. 

%Fig%%%%%%%%%%%%%%%%%%%%%%%%%%%%%%%%%%%%%%%%%%%%%%%%%%%%%%%%%%%%%%%%%%%%%%%%%%%%%%%%%%%%%%%%%%%%%%%%%%%%%%%%%%
% Figure 6
%Fig%%%%%%%%%%%%%%%%%%%%%%%%%%%%%%%%%%%%%%%%%%%%%%%%%%%%%%%%%%%%%%%%%%%%%%%%%%%%%%%%%%%%%%%%%%%%%%%%%%%%%%%%%%

\begin{figure}
\centerline{\includegraphics[width=0.9\textwidth,clip=,angle=0]{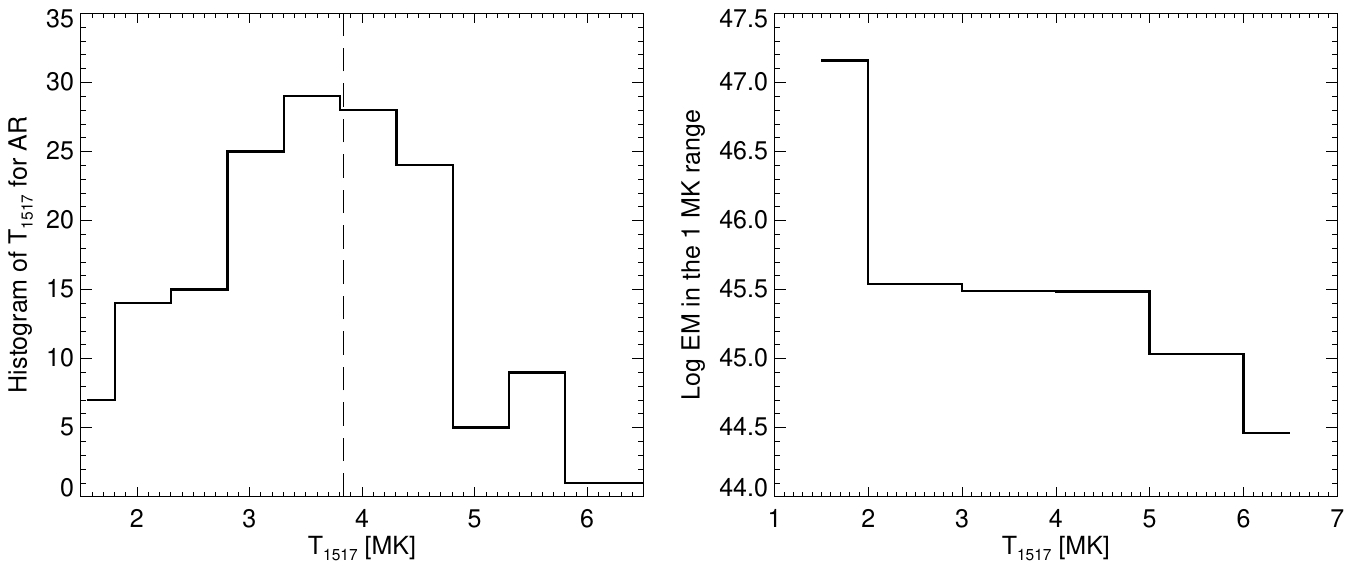}}
\caption{Left: Distribution of temperatures for the 158 nonflaring active region spectra. The dashed vertical line is shown for the average temperature 3.84~MK. Right: Logarithmic plot of emission measure distribution (units of cm$^{-3}$, 1-MK temperature intervals) within the 14\arcsec\ x 14\arcsec\ field of view. \label{fig:fig7}}
\end{figure}

The distribution of temperatures for the nonflaring active regions studied is an important property of coronal active regions generally, and may be compared with other studies from, e.g., extreme ultraviolet spectra and X-ray instruments on the Hinode spacecraft \citep{tes11}. Figure~\ref{fig:fig7} shows the distribution of measured temperatures $T_{1517}$ and corresponding volume emission measures $EM = N_e^2 V$ in cm$^{-3}$ ($N_e =$ electron density in cm$^{-3}$ and $V = $ emitting volume in cm$^3$) for the active regions studied here. The average temperature was found to be 3.84~MK (indicated in the plot), hence the temperature used for the CHIANTI spectrum in Figure~\ref{fig:fig3}. The distribution sharply decreases for $T_{1517} > 5.5$~MK. There is a remarkable similarity between these distributions and that of the nonflaring active region studied by \cite{par73}. The emission measures in Figure~\ref{fig:fig7} are for active region cores, or for the FCS those within the FCS collimator (14\arcsec\ $\times$ 14\arcsec). If the emission is uniform over a cube with side 14\arcsec\ ($1.02 \times 10^9$~cm), equal to a volume $1.05 \times 10^{27}$~cm$^3$, the corresponding electron density is approximately $10^{10}$~cm$^{-3}$. This is more than estimates from instruments on the Hinode spacecraft (e.g. \cite{tes11}), although this may be explained by the fact that the Hinode observations referred to were not in active region cores but in loop structures away from the center of the active region. The emission observed by FCS probably arises from a few unresolved emission centers with smaller volumes and correspondingly even higher electron density. 

Figure~\ref{fig:fig3} comparing the average FCS channel~1 spectra from nonflaring  active regions with a CHIANTI spectrum with a temperature of 3.8~MK shows a particularly large disagreement for the \ion{Fe}{17} 3C line at 15.015~\AA. This disagreement cannot be easily explained by instrumental effects such as variations in instrumental sensitivity, and consequently in the analyses of \cite{sch92}, \cite{phi96a}, and \cite{phi97} the authors suggested that the 3C line intensity is reduced by resonant scattering, citing large values then current of the line's oscillator strength. An apparent correlation of the amount of line 3C's reduction with central meridian distance of the active region seemed to support this suggestion. However, more recent EBIT measurements \citep{bro06,kuh22} indicate a much reduced oscillator strength for this line, so there is now no need to invoke optical thickness for the 3C line. The explanation for the erroneous conclusions from the earlier works may be underestimations of time variations in the course of spectral scans. \cite{bro01}'s EBIT spectra also show that blending with the 3D line with an intense satellite feature (called ``C'') accounts for the apparent time variation of the intensity ratio of the 3C and 3D lines. The EBIT spectra also show other satellites apparent in FCS spectra and which are shown in Figure~\ref{fig:fig8} discussed below.

Using both nonflaring active region and flare spectra, we found that the intensity of the 3C line is closely proportional to that of the 3F line, which has no significant blending with other lines.  This is shown in Figure~\ref{fig:fig9}, with observed total emission (units of photon counts s$^{-1}$) for each line. In this plot, both nonflaring active-region and flare points have been included.  There are very few outliers from the proportionality line, the occurrence of which would be expected for emission near the solar limb when line 3C if optically thick might have reduced emission. Thus, Figure~\ref{fig:fig9} strongly suggests that the 3C \ion{Fe}{17} line is not optically thick, in agreement with the EBIT findings. 

With the measured temperatures $T_{1517}$ and corresponding emission measures $EM = N_e^2 V$ (cm$^{-3}$) for the active regions, the fluxes divided by the emission measure (units of $10^{46}$~cm$^{-3}$) can be derived for the \ion{Fe}{17} lines and can be compared with contribution functions (as defined in the usual way, e.g. \cite{phi08}) from theory. This is shown in Figure~\ref{fig:fig10} for lines 3C, 3D, 3F, and the sum of lines 3G and 3H, with contribution functions from CHIANTI v.~11 (red lines). These are from collision strengths calculated by the $R$-matrix formalism by \cite{lia10}. Clearly, over the temperature range illustrated, there is excellent agreement for the relatively unblended lines 3F and the sum of 3G and 3H. There is, however, departure at high-temperatures ($>4$~MK) for lines 3C and 3D, and at low temperatures $<2.5$~MK for line 3D. The latter is explained by the presence of the \ion{Fe}{16} dielectronic satellite ``C'' in EBIT spectra \cite{bro01} which has almost identical wavelength to \ion{Fe}{17} line 3D. The discrepancy at higher temperatures with lines 3C and 3D may be due to a shortcoming in the collision strength calculations. Although this is unlikely to affect most of the FCS nonflaring active region spectra that have lower temperatures, it may point to a problem in the collision strength calculations and be a contributing factor in the differences between the observed and calculated spectra in Figure~\ref{fig:fig3}. We note that there is a small difference in the \cite{lia10} calculations of the oscillator strength 3C/3D ratio (4.07) and the later calculations (3.55) and measurements (3.51) of \cite{kuh22}, although this difference is unlikely to explain the 3C and 3D discrepancies indicated in Figure~\ref{fig:fig10}. The slightly worsening disagreement of the 3C and 3D curves at higher temperatures in Figure~\ref{fig:fig10} is supported by the EBIT measurements of \cite{lam00}, indicating decreasing 3C/3D ratios at higher electron beam energies (see also \cite{bro01} although their measurements used a crystal spectrometer and are more sensitive to EBIT polarization issues).

%Fig%%%%%%%%%%%%%%%%%%%%%%%%%%%%%%%%%%%%%%%%%%%%%%%%%%%%%%%%%%%%%%%%%%%%%%%%%%%%%%%%%%%%%%%%%%%%%%%%%%%%%%%%%%
% Figure 7
%Fig%%%%%%%%%%%%%%%%%%%%%%%%%%%%%%%%%%%%%%%%%%%%%%%%%%%%%%%%%%%%%%%%%%%%%%%%%%%%%%%%%%%%%%%%%%%%%%%%%%%%%%%%%%

% Subsection 3.2
\subsection{Fe XVI satellites}
\label{FeXVI_sats}

\begin{figure}
\centerline{\includegraphics[width=0.8\textwidth,clip=,angle=0]{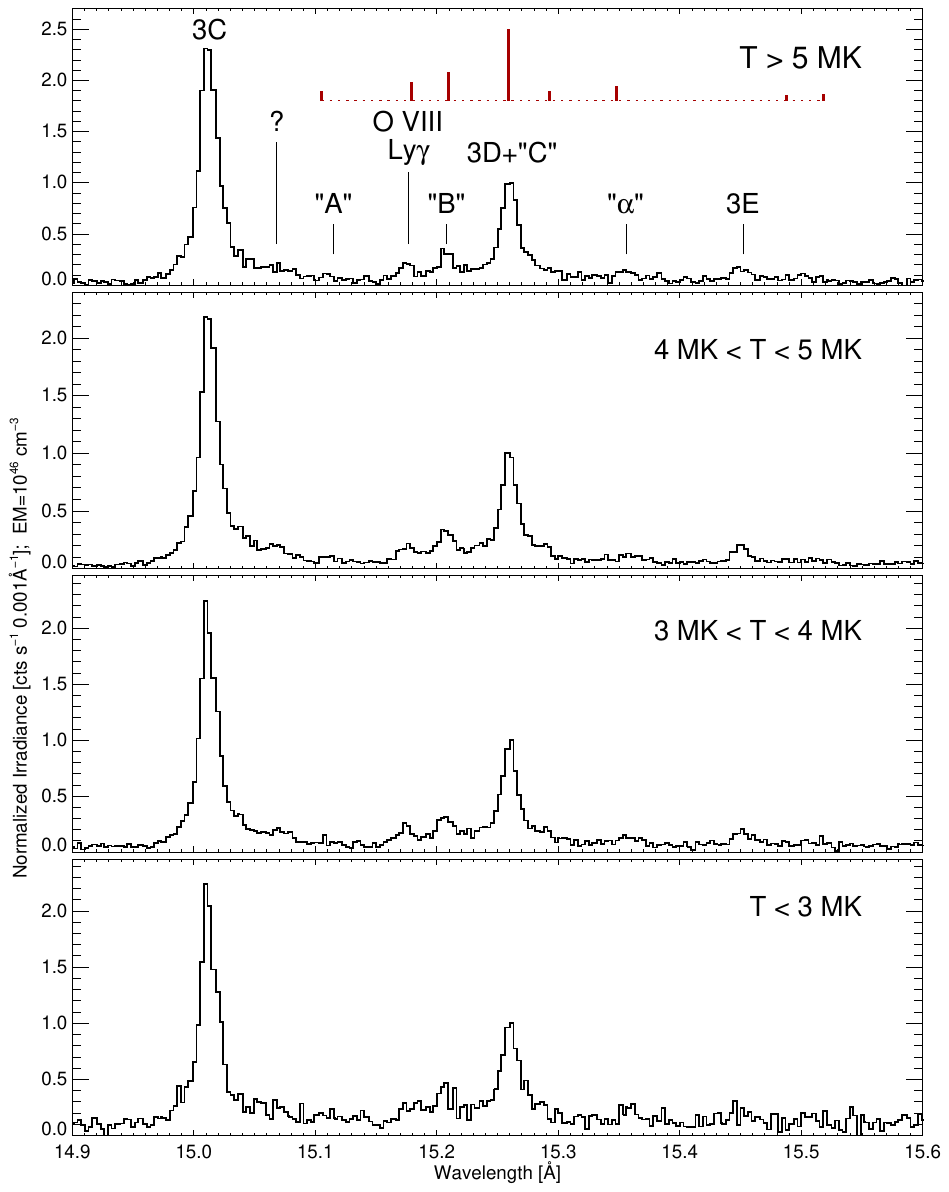}}
\caption{Left: Averaged FCS channel 1 spectra over the wavelength range 14.9~\AA\ to 15.6~\AA\ in the four temperature bands indicated, lowest temperature at the bottom. The vertical scale units are photon counts~s$^{-1}$ in wavelength intervals of 0.001~\AA\ and normalized to a volume emission measure of $10^{46}$~cm$^{-3}$. These low nonflaring active region temperatures show the \ion{Fe}{16} dielectronic satellites observed with EBIT by \cite{bro01,bro06} to best advantage. The letter notation in these works is indicated in the top panel, as are the wavelengths and intensities of the strongest \ion{Fe}{16} satellites from \cite{delz24}. The line at 15.1762~\AA\ is not an \ion{Fe}{16} satellite but is due to \ion{O}{8} Ly-$\gamma$. \label{fig:fig8}}
\end{figure}

\begin{figure}
\centerline{\includegraphics[width=0.44\textwidth,clip=,angle=0]{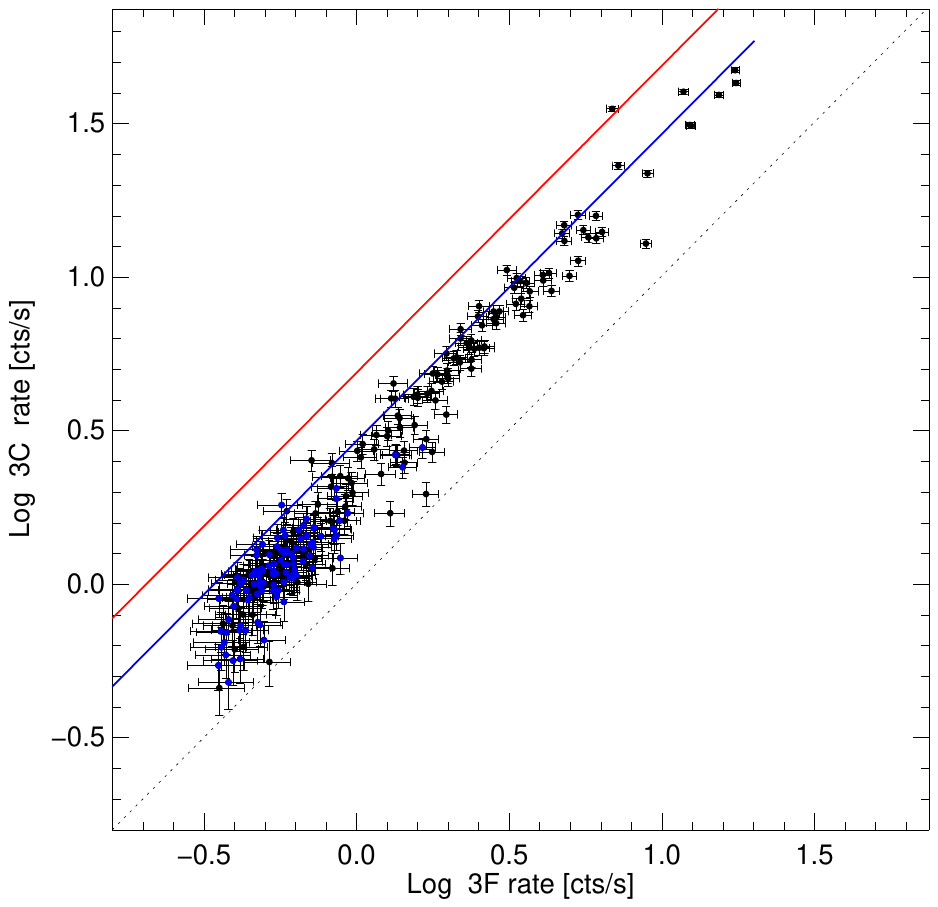}}
\caption{Logarithmic plot of intensity of \ion{Fe}{17} line 3C against line 3F (units: photon counts s$^{-1}$) for nonflaring active regions (blue points) and flares. The error bars reflect statistical uncertainties. The dotted line shows the intensity of line 3C equal to that of line 3F, the blue line 3C twice the line 3F intensity, the red line three times the line 3F intensity, illustrating the proportionality of the intensities of the two lines. 
\label{fig:fig9}}
\end{figure}

\begin{figure}
\centerline{\includegraphics[width=0.5\textwidth,clip=,angle=90]{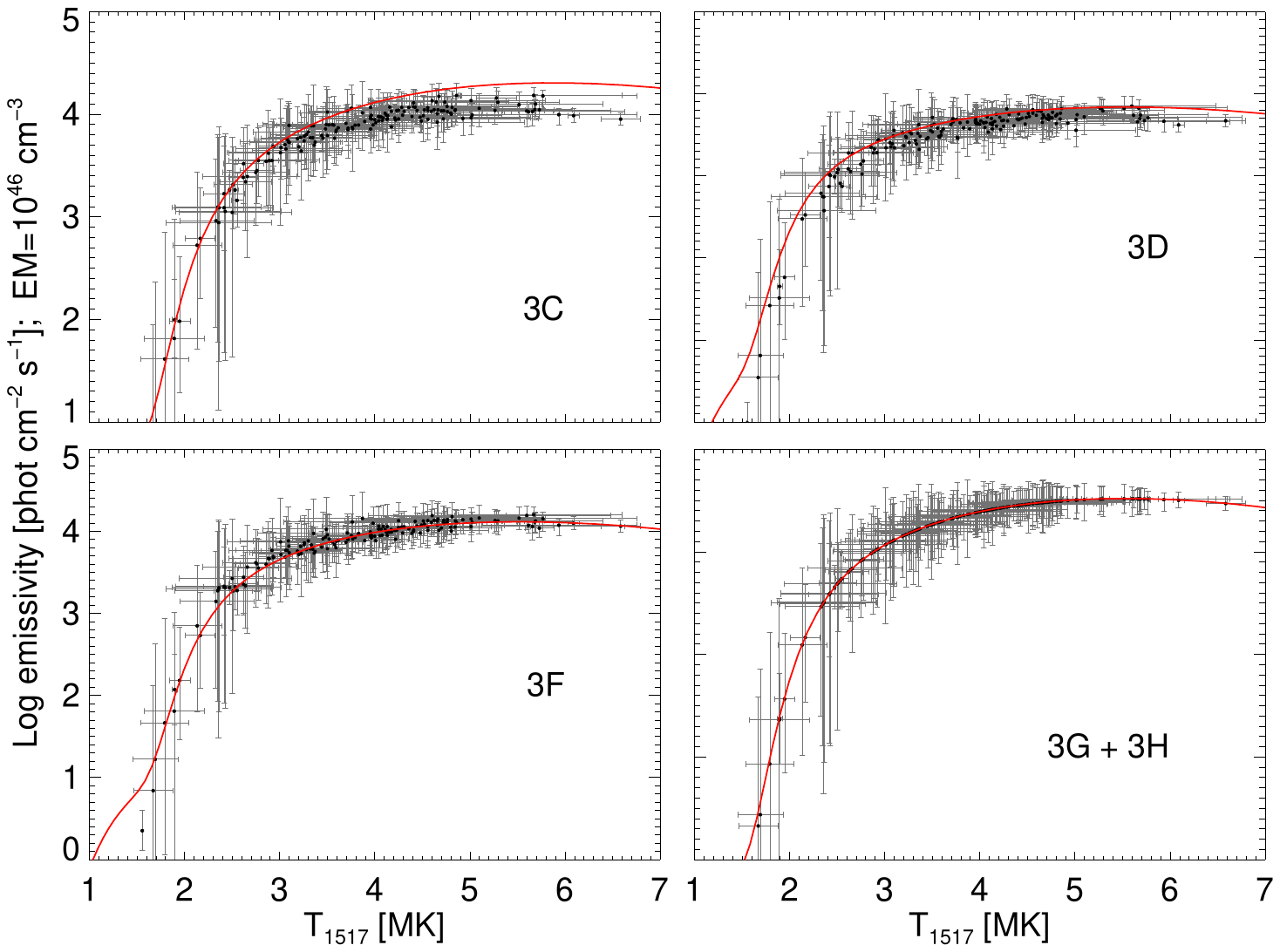}}
\caption{Observed line emission shown as points with error bars for principal \ion{Fe}{17} lines (photon counts s$^{-1}$) normalized for a volume emission measure $EM = 10^{46}$~cm$^{-3}$ with pre-launch values for the FCS channel~1 effective areas. Top left: line 3C (15.0130~\AA\ observed wavelength, Table~\ref{tab:line_ids}). Top right: line 3D (15.2611~\AA). Bottom left: line 3F (16.7768~\AA). Bottom right: sum of lines 3G (17.0504~\AA) and 3H (17.0957~\AA).  In each plot, the observed points are compared with CHIANTI v.~11 calculations (red line) with additions for \ion{Fe}{16} satellites (G. Del Zanna, private communication). 
\label{fig:fig10}}
\end{figure}

As well as the \ion{Fe}{17} lines, a number of \ion{Fe}{16} satellites are relatively strong which we now discuss. Calculations of the \ion{Fe}{16} satellite wavelengths and intensity factors (generally called $F_2$, units of s$^{-1}$: see \cite{gab72}) have been given by \cite{bei14} and more recently by \cite{delz24} (see their Table~7). In the following, we have identified the strongest satellites seen in the FCS active region spectra (where the relatively low temperatures enable the satellites to be seen most clearly) with the \cite{delz24} calculations. We have also identified \ion{Fe}{16} satellites with features observed with the Lawrence Livermore National Laboratory (LLNL) EBIT \cite{bro01}, labeled by them as ``A'', ``B'', and ``$\alpha$''. (We denote these satellites with quotation marks to distinguish them from the \ion{Fe}{17} lines 3A -- 3H.) Figure~\ref{fig:fig8} shows FCS spectra in the range 14.9~\AA\ to 15.6~\AA\ normalized to an emission measure of $10^{46}$~cm$^{-1}$ in four intervals of $T_{1517}$ showing the EBIT satellites to best advantage.    The relatively strong satellite ``B'' at 15.2080~\AA\ appears to be due to the \ion{Fe}{16} line with transition $2s^2 2p^6\,3s ^2S_{1/2} - 2s^2 2p^5 3s 3d \,\,^2P_{1/2}$, as was noted by \cite{bri06}. The ``A'' line at 15.1100~\AA, which is very weak both in the FCS and EBIT spectra, is probably the \ion{Fe}{16} $2s^2 2p^6\,3s ^2S_{1/2} - 2s^2 2p^5 3s 3d \,^2P_{3/2}$ line. For the \cite{bro01} line ``C'', we note that the EBIT spectrum of \cite{kuh22} indicates an energy difference of only 0.08~eV from the \ion{Fe}{16} line 3D, corresponding to a wavelength difference of 0.0015~\AA, far smaller than the wavelength resolution of any solar X-ray spectrometer. From \cite{delz24}, this \ion{Fe}{16} line has a transition $2s^2 2p^6\,3s\, ^2S_{1/2} - 2s^2 2p^5 3s3d\, ^2D_{3/2}$. Finally, we identify \cite{bro01}'s line ``$\alpha$'' with the \ion{Fe}{16} line with transition $2s^2 2p^6\,3d\, ^2D_{3/2} - 2s^2 2p^5 3d^2\, ^2F_{5/2}$, although \cite{bro01} identify this line, seen as a fairly strong feature in both their EBIT spectra as well as our FCS spectra, as a \ion{Fe}{15} line. This is still possibly correct, but for FCS spectra it is more likely due to an \ion{Fe}{16} transition as it is seen in FCS spectra at temperatures as high as 5~MK when the ionization fraction of Fe$^{+14}$ ions is small. In Figure~\ref{fig:fig8} (top panel), we mark the wavelengths and intensities of the strongest \ion{Fe}{16} satellites from \cite{delz24}. Table~\ref{tab:line_ids} gives these identifications together with measured FCS channel~1 wavelengths. 

The presence of many unresolved, very weak satellites, probably due to \ion{Fe}{16}, is indicated in Figure~\ref{fig:fig8} as a ``shoulder'' of emission on the long-wavelength side of line 3C. 

Comparison of calculated \ion{Fe}{16} wavelengths from various authors shows a fairly considerable uncertainty, especially for weaker satellites not observed as separate features. However, we may still safely use the group of satellites to the longward of line 3C and 3D ($F_{15{\rm sat}}$) and the group on the longward side of lines 3G and 3H ($F_{17{\rm sat}}$) in the ratio with the \ion{Fe}{17} lines 3F and 3G+3H respectively to determine the temperature $T_{1517}$ as the precise wavelengths are not critical to this ratio. 

%Fig%%%%%%%%%%%%%%%%%%%%%%%%%%%%%%%%%%%%%%%%%%%%%%%%%%%%%%%%%%%%%%%%%%%%%%%%%%%%%%%%%%%%%%%%%%%%%%%%%%%%%%%%%%
% Figure 8
%Fig%%%%%%%%%%%%%%%%%%%%%%%%%%%%%%%%%%%%%%%%%%%%%%%%%%%%%%%%%%%%%%%%%%%%%%%%%%%%%%%%%%%%%%%%%%%%%%%%%%%%%%%%%%

%Fig%%%%%%%%%%%%%%%%%%%%%%%%%%%%%%%%%%%%%%%%%%%%%%%%%%%%%%%%%%%%%%%%%%%%%%%%%%%%%%%%%%%%%%%%%%%%%%%%%%%%%%%%%%
% Figure 9
%Fig%%%%%%%%%%%%%%%%%%%%%%%%%%%%%%%%%%%%%%%%%%%%%%%%%%%%%%%%%%%%%%%%%%%%%%%%%%%%%%%%%%%%%%%%%%%%%%%%%%%%%%%%%%

%Fig%%%%%%%%%%%%%%%%%%%%%%%%%%%%%%%%%%%%%%%%%%%%%%%%%%%%%%%%%%%%%%%%%%%%%%%%%%%%%%%%%%%%%%%%%%%%%%%%%%%%%%%%%%
% Figure 10
%Fig%%%%%%%%%%%%%%%%%%%%%%%%%%%%%%%%%%%%%%%%%%%%%%%%%%%%%%%%%%%%%%%%%%%%%%%%%%%%%%%%%%%%%%%%%%%%%%%%%%%%%%%%%%

% Subsection 3.3
\subsection{Fe XVII G and H lines: Density lower limit}
\label{GandH_lines}

The \ion{Fe}{17} 17~\AA\ lines have not attracted as much attention as the 15~\AA\ lines but they are of great interest owing to the highly forbidden nature of the magnetic quadrupole 17.10~\AA\ (3H) line. This leads to the 17~\AA\ lines having a density sensitivity, as has been discussed by \cite{bha85} for a number of Ne-like ions. For comparatively low-$Z$ ions such as \ion{Si}{5} and \ion{Ar}{9}, the intensity ratio 3G/3H remains constant over a range of electron densities $N_e$ up to a critical value beyond which the intensity of the forbidden (3H) line decreases sharply with respect to the intercombination 3G line. The critical values are $N_e = 10^8$ cm$^{-3}$ for Si and $10^{12}$ cm$^{-3}$ for Ti. For \ion{Fe}{17}, the critical density, although not specifically calculated by \cite{bha85}, is presumably larger than $10^{12}$~cm$^{-3}$, implying a constant value for the 3G/3H intensity ratio in the low-density limit for nonflaring active region spectra. This  was found to be the case for the nonflaring active region spectra in this study, with the average 3G/3H ratio equal to 1.068 (see Table~\ref{tab:line_ids}). The \ion{Ni}{19} lines 13.553~\AA\ and 13.698~\AA\ have a similar ratio but are much weaker. There is a need of revised calculations of line ratios based on improved atomic data to see whether agreement with the FCS and other observations can be achieved. 

Observations of very high densities ($\sim 3 \times 10^{12}$~cm$^{-3}$) have been noted from FCS spectra taken at or near the impulsive stage of a flare in 1985 \citep{phi96b}. It is possible that the density diagnostic described by \cite{bha85} for \ion{Fe}{17} lines would be applicable for flares in their impulsive stages. 

% Section 3.4
\subsection{Individual Active Regions}
\label{Indiv_ARs}

\begin{figure}
\centerline{\includegraphics[width=0.80\textwidth,clip=,angle=0]{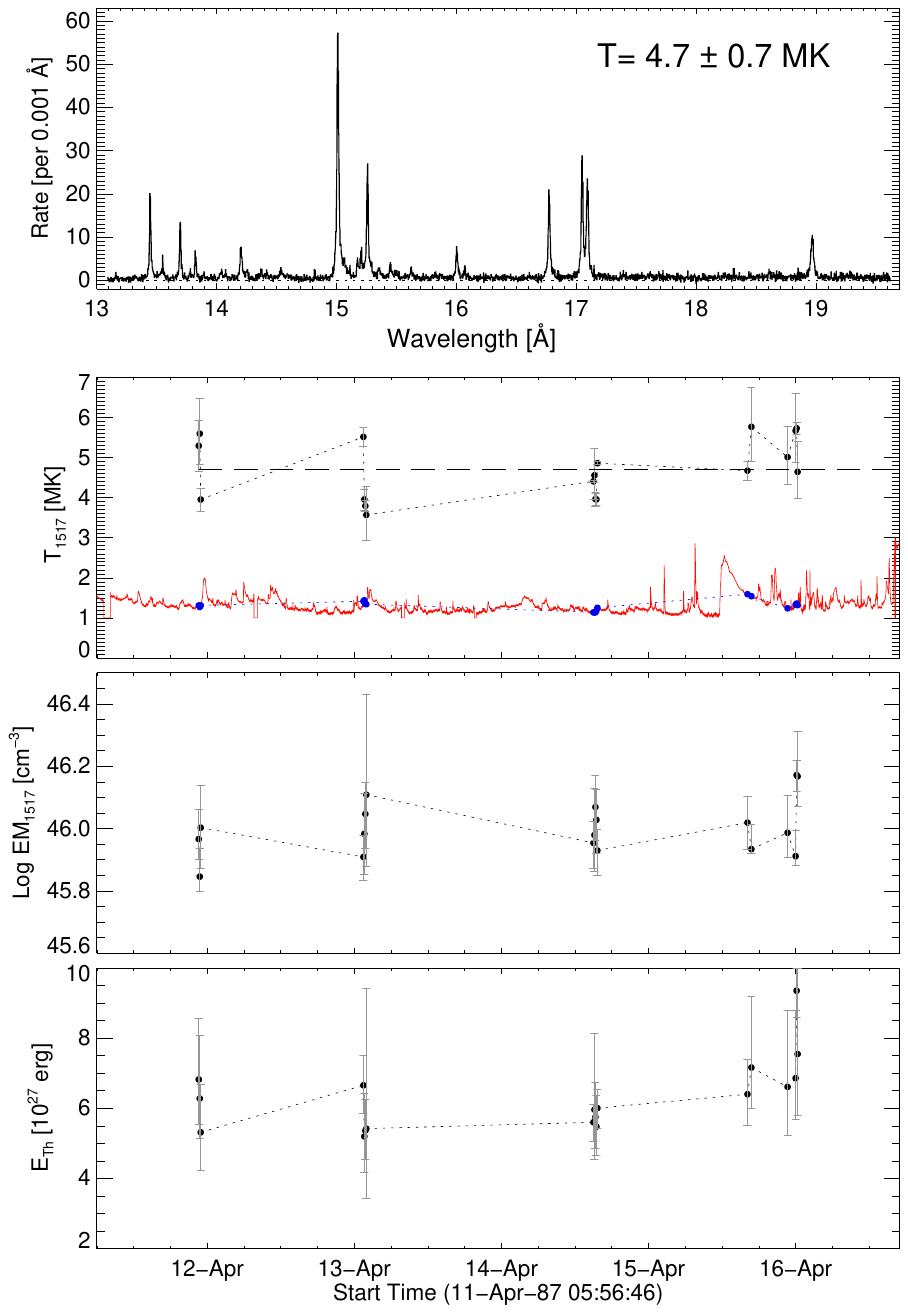}}
\caption{Average spectrum for 18 FCS spectra taken of AR4787 (top panel) and time evolution of the 18 measured temperatures $T_{1517}$ in AR4787 (second from top), emission measures $EM = N_e^2 V$, and thermal energies (defined in the text) within a volume equal to (14\arcsec)$^3$ ($= 1.05 \times 10^{27}$~cm$^3$) relating to the FCS field of view.  In red (second panel) the logarithm of the GOES 1-8~\AA~flux on a relative scale is shown (blue points are for the times of the FCS spectra). 
\label{fig:fig11}}
\end{figure}

Two active regions were particularly well observed by the FCS in 1987, known by their NOAA designations: AR4787 (April) and AR4901 (December). A series of 18 FCS scans made between 1987 April~11 and April~17 from AR4787 in a nonflaring state allowed a sequence of temperatures $T_{1517}$ and volume emission measures $EM_{1517}$, as well as estimates of a quantity related to total thermal energy, defined by $E_{Th} = 3kT_{1517} N_e V = 3kT \sqrt(EM_{1517}) \times \sqrt{V} $, where $k$ is Boltzmann's constant and $V$ the volume viewed by the FCS, assumed to be a cube with side 14\arcsec\ or $V = 1.05 \times 10^{27}$~cm$^3$. There is little change over the  seven-day period, but there are indications of a slight increase as the region approached the west limb on April~19, as indicated by the increase in GOES emission. This is indicated more by the time variations of $E_{Th}$ than the temperature and emission measure. The observations are listed in Table~\ref{tab:AR4787} and summarized in Figure~\ref{fig:fig11}. (Note that active region AR4787 is given by \cite{delz14} in their analysis as AR4780.)

%Fig%%%%%%%%%%%%%%%%%%%%%%%%%%%%%%%%%%%%%%%%%%%%%%%%%%%%%%%%%%%%%%%%%%%%%%%%%%%%%%%%%%%%%%%%%%%%%%%%%%%%%%%%%%
% Figure 11

%Fig%%%%%%%%%%%%%%%%%%%%%%%%%%%%%%%%%%%%%%%%%%%%%%%%%%%%%%%%%%%%%%%%%%%%%%%%%%%%%%%%%%%%%%%%%%%%%%%%%

\begin{deluxetable*}{lcccccc}
%\tabletypesize{\tiny}
\tabletypesize{\small}
\tablecaption{FCS Nonflaring Spectral Scans from AR4787/4790. \label{tab:AR4787} }
\tablewidth{0pt}
\tablehead{
\colhead{Sp. no.} & \colhead{Date}  & \colhead{Scan times (UT)} & Estimated & Estimated & \colhead{GOES 1--8~\AA\ } & \colhead{Heliographic} \\ 
\colhead{}&{}& (start - end) & Temp. (MK) & EM ($10^{46}$~cm$^{-3}$) & level & coords.\\}
\startdata 
1&    1987-Apr-11& 22:32:06 - 22:41:27&   5.3&  0.9& B2.0&  S33E24\\
2&    1987-Apr-11& 22:41:45 - 22:51:06&   5.6&  0.7& B1.9&  S33E24\\
3&    1987-Apr-11& 22:51:25 - 23:00:46&   4.0&  1.0& B2.0&  S33E24\\
4&    1987-Apr-13& 01:24:01 - 01:33:22&   5.5&  0.8& B2.6&  S34W05\\
5&    1987-Apr-13& 01:33:41 - 01:43:02&   4.0&  1.0& B2.8&  S34W05\\
6&    1987-Apr-13& 01:43:20 - 01:52:41&   3.8&  1.1& B2.2&  S34W05\\
7&    1987-Apr-13& 01:53:00 - 02:02:21&   3.6&  1.3& B2.2&  S34W05\\
8&    1987-Apr-14& 14:56:22 - 15:05:43&   4.4&  0.9& B1.4&  S34W15\\
9&    1987-Apr-14& 15:06:02 - 15:15:23&   4.6&  1.0& B1.4&  S34W15\\
10&   1987-Apr-14& 15:15:42 - 15:25:03&   4.0&  1.2& B1.4&  S34W15\\
11&   1987-Apr-14& 15:25:22 - 15:34:43&   4.0&  1.1& B1.5&  S34W15\\
12&   1987-Apr-14& 15:35:01 - 15:44:22&   4.9&  0.9& B1.8&  S34W15\\
13&   1987-Apr-15& 16:04:16 - 16:13:37&   4.7&  1.1& B3.9&  S31W37\\
14&   1987-Apr-15& 16:42:55 - 16:52:16&   5.8&  0.9& B3.5&  S31W37\\
15&   1987-Apr-15& 22:34:08 - 22:48:19&   5.0&  1.0& B1.7&  S31W40\\
16&   1987-Apr-15& 23:55:30 - 00:04:51&   5.7&  0.8& B2.1&  S31W40\\
17&   1987-Apr-16& 00:05:10 - 00:14:31&   5.7&  1.5& B2.3&  S31W40\\
18&   1987-Apr-16& 00:14:50 - 00:24:11&   4.6&  1.5& B2.1&  S31W40\\
\enddata
\end{deluxetable*}

% Section 4
\section{Conclusions}
\label{Conclusions}

Archived data from the Flat Crystal Spectrometer on Solar Maximum Mission, operating from 1980 to 1989, have been analyzed in this study largely on the basis that the channel~1 spectra, covering the range 13~\AA\ to 22~\AA, have a resolution unsurpassed by any instrument flown since the 1980s. The data include 165 nonflaring active region spectra from archived data. This spectral range includes  \ion{Fe}{17} line spectra so allowing their line intensities to be compared with theory from the CHIANTI atomic data package and laboratory data. A large number of \ion{Fe}{16} dielectronic satellites occur in the neighborhood of the 15~\AA\ lines (3C and 3D) in particular, the most prominent of which can be compared with spectra from the EBIT laboratory device, and the satellites observed by \cite{bro01} can be identified using the recent work of \cite{delz24}. As the emission from nonflaring active regions has a relatively low temperature ($< 6$~MK), the usual temperature diagnostics based on X-ray line ratios cannot be used, so instead we took advantage of the inverse temperature dependence of the ratio of \ion{Fe}{16} satellites on the long-wavelength side of \ion{Fe}{17} lines at 15~\AA\ and 17~\AA\ to \ion{Fe}{17} unblended lines. This enabled us to arrange the spectra in order of temperature, showing that the \ion{Fe}{17} lines are emitted over the range from less than 3~MK to more than 6~MK. Comparison with CHIANTI spectra shows that \ion{Fe}{17} lines 3C and 3D are underintense but we show that this cannot be due to resonant scattering as was mistakenly concluded in previous analyses. The contribution functions for these lines, unlike the unblended lines 3F, 3G, and 3H, show some departure from the CHIANTI theory at high temperatures. We believe that revised calculations of the collisional excitation rates may be necessary and may solve the problem of the underintensities of lines 3C and 3D. The intensity ratio 3G to 3H is in theory density-dependent but for the Fe lines this is only true for very high densities. Departures from the 3G/3H ratio from the low-density limit may, however, be observed for the very high densities discerned at the impulsive stage of flares, which future spectrometers with high enough resolution might take advantage of. 

The background emission in FCS spectra, arising from fluorescence mainly from crystals near to the KAP crystal of channel~1, could be avoided with a simple instrumental redesign such as small collimators on each crystal. This was recognized for the proposed ChemiX (a Bragg crystal spectrometer for the \textit{Interhelioprobe}  satellite (see \cite{sia16}) and a projected spectrometer called BRAXIS for a possible medium-class NASA solar mission (J. Sylwester, lead scientist). 

In future work, we will concentrate on solar flare spectra from the FCS, in particular electron densities and properties of the flaring plasma. 

\begin{acknowledgments}

This research was funded by the National Science Centre, Poland, Grants Nos. 2020/39/B/ST9/01591 and 2023/49/B/ST9/02409. The CHIANTI atomic database and code is a collaborative project involving George Mason University, the University of Michigan (USA), and the University of Cambridge (UK). 

We acknowledge help given by Dr Chintan Shah with his EBIT observations, to Dr Giulio Del~Zanna with his recent work on satellite line emission, and to Professor Helen Mason at Cambridge University, U.K. We also thank the anonymous referee for clarifying some issues.
\end{acknowledgments}

\vspace{5mm}
\facilities{Solar Maximum Mission (FCS)}
\software{SolarSoft Interactive Data Language \citep{fre98}, {\sc chianti} \citep{delz15}. }

\bibliography{RESIK}
\bibliographystyle{aasjournal}

\end{document}